\documentclass[12pt,letterpaper]{article}
\usepackage{osajnl}
\usepackage{overcite,hyperref}       

\begin{document}

\title{Mode areas and field energy distribution in honeycomb photonic bandgap fibers}
\author{Jesper L{\ae}gsgaard,$^1$ Niels Asger Mortensen,$^2$ and Anders Bjarklev$^1$}
\affiliation{$^1$Center for Communication, Optics and Materials (COM), Technical Univ. of Denmark Bldg. 345v,
 DK-2800 Kgs. Lyngby, Denmark,\\ 
$^2$Crystal Fibre A/S, Blokken 84, DK-3460 Birker{\o}d, Denmark}
%
%
\begin{abstract}

The field energy distributions and effective mode areas of silica-based photonic
bandgap fibers with a honeycomb airhole structure in the cladding and an extra 
airhole defining the core are investigated. We present a generalization of 
the common effective area definition, suitable for the problem at hand, and 
compare the results for the photonic bandgap fibers with those of index-guiding
microstructured fibers.  While the majority of the field energy in the 
honeycomb photonic bandgap fibers is found to reside in the silica, a 
substantial fraction (up to $\sim$ 30 \%) can be located in the
airholes. This property may show such fibers particularly interesting for
sensor applications, especially those based on nonlinear effects or interaction
with other structures (e.g. Bragg gratings) in the glass.  

\end{abstract}
\maketitle
%
%
\newcommand{\be}{\begin{equation}}
\newcommand{\ee}{\end{equation}}
\newcommand{\bea}{\begin{eqnarray}}
\newcommand{\eea}{\end{eqnarray}}
\newcommand{\eps}{\varepsilon}
\newcommand{\om}{\omega}
\newcommand{\aeff}{$A_{\rm eff}$ }
\newcommand{\aefft}{$\tilde{A}_{\rm eff}$ }
\newcommand{\maeff}{A_{\rm eff} }
\newcommand{\maefft}{\tilde{A}_{\rm eff} }
\newcommand{\dloge}{\frac{\partial\ln\eps}{\partial\om}}
\newcommand{\dcol}{$\frac{d_c}{\Lambda}$ }
\newcommand{\dclol}{$\frac{d_{cl}}{\Lambda}$ }
\newcommand{\dcolns}{$\frac{d_c}{\Lambda}$}
\newcommand{\dclolns}{$\frac{d_{cl}}{\Lambda}$}

\section{Introduction}
 
  Photonic crystal fibers (PCFs), which guide light in a single-material 
structure by coherent scattering from an array of $\mu$m-sized airholes 
(for recent reviews we refer to Refs.~\onlinecite{knight2002,birks2001} and 
references therein), have in recent years emerged as an attractive alternative 
to conventional optical fibers within the area of nonlinear fiber 
devices.\cite{broderick-1999,ranka-2000}  The advantages of the PCFs are 
firstly that very small mode areas can be obtained due to
the large refractive index contrast between silica and air, leading to high
nonlinearity coefficients. Secondly, the PCFs allow for a more flexible 
tailoring of the dispersion properties, which are crucial for many applications.
PCFs with zero-dispersion wavelengths ranging from 565 to 1550 nm and high
nonlinearity coefficients have been demonstrated.\cite{knight2000,kph-ofc02} 

   The highly nonlinear PCFs fabricated today are of the index-guiding type\cite{knight1996}, in which
a missing hole in a triangular lattice of airholes defines a high-index core,
which guides light by total internal reflection. Fibers guiding light in large
hollow cores by means of the photonic band gap (PBG) effect have also been 
demonstrated \cite{cregan1999}, with the intention of obtaining very low losses and 
nonlinearities. However, an alternative PBG fiber design in which a honeycomb
airhole lattice is modified by addition of an airhole to form a low-index core region \cite{broeng1998} has until now not been investigated thoroughly although its 
practical feasibility was demonstrated as early as 1998.\cite{knight1998} The design is shown
in Fig.~\ref{fig1}, and can be characterised by three parameters: The physical distance
between nearest-neighbor airholes (commonly denoted the pitch, or $\Lambda$),
the diameter of the cladding holes, $d_{cl}$, and the diameter of the hole forming
the core defect, $d_c$. These fibers guide the majority of the light in silica,
just as the conventional index-guiding PCFs, and may, therefore, constitute an
alternative way of fabricating highly nonlinear fibers. The purpose of the
present work is to investigate the design depicted in Fig.~\ref{fig1} with respect
to field energy distribution and nonlinear coefficients. It will be shown
that nonlinearities comparable to those obtained in index-guiding PCFs can
be achieved in the honeycomb PBG fibers, while at the same time, a substantial
fraction of the field energy may be pushed into the airholes. This may make
these fibers particularly interesting for applications as sensing
devices.\cite{monro-evf-2001,hoo-evf}

    The rest of the paper is organized as follows: In Sect. 2, we describe
our theoretical methods, and derive a generalized formula for the fiber
nonlinearity coefficient, in terms of an effective area, which is valid for
all field distributions, including those where a substantial part of the field energy resides in air. In
Sect. 3, we present and discuss our numerical results for some selected 
honeycomb designs, and compare them to results for index-guiding PCFs. Finally, Sect. 4 summarizes our conclusions.

\section{Theoretical approach}

 We consider a structure, which is uniform along the $z$-axis while structured
in the $x$-$y$ plane. The magnetic field vector, {\bf H}, may then be written:
\be
{\bf H}({\bf r})=e^{i(\om t-\beta z)}{\bf H}(x,y),
\ee
for a monochromatic wave. The fundamental equation for {\bf H} is:
\bea
\frac{\om^2}{c^2}=\frac{\langle{\bf H}, {\bf \Theta H}\rangle}{\langle {\bf H},{\bf H}\rangle},\label{feq}\\
{\bf \Theta H}=\nabla\times\frac{1}{\eps_r({\bf r})}\nabla\times{\bf H}
\eea
where $\eps({\bf r})=\eps_0\eps_r({\bf r})$ is the dielectric function. For a fixed propagation
constant, $\beta$, this equation may be solved for $\om$, which can then be
regarded as a function of $\beta$ and $\eps$({\bf r}). From the magnetic field
vector, the corresponding electric field is straightforwardly obtained using
Amperes law (SI units are used throughout the paper):

\be
 \nabla\times {\bf H} = i\om\eps {\bf E}
\label{lexampere}
\ee

    The effective area is a concept originating in the theory of third-order
nonlinearity effects in optical waveguides\cite{agrawal}. In a homogeneous 
material, such as amorphous silica, the third-order part of the nonlinear 
susceptibility gives rise to an amplitude dependent shift in the material 
refractive index:

\be
n=n_0 + \Delta n = n_1 + n_2\mid {\bf E}\mid^2
\ee
where $n_1$ is the refractive index of silica in the limit of zero field, and
$n_2$ is a nonlinear coefficient related to the nonlinear-susceptibility tensor,
$\chi^{(3)}$, through:\cite{agrawal}

\be
n_2=\frac{3}{8n_1}\mathrm{Re}\chi^{(3)}_{xxxx}.
\ee

In an optical fiber, the change in material index leads to a corresponding 
change in the effective index ($n_{\rm eff}=\frac{\beta}{\omega}$).
In first-order perturbation theory, the mode-field
distribution can be considered unchanged by the index perturbation, and the
change in $\om$ for a (Kerr-induced) $\Delta\eps$  for fixed $\beta$ is from Eq.
(\ref{feq}) found to be:

\be
\Delta\om=-\frac{\om \eps_0c^2\langle {\bf E},\Delta\eps_r {\bf E}\rangle}{2
\langle {\bf H,H}\rangle}. 
\label{dwdepseq}
\ee

Usually, the experimental situation is that light is launched at a fixed 
frequency $\om_0$ and the Kerr-induced change in refractive index effects a
change in the propagation constant $\beta$ from $\beta_0$, say, to 
$\beta_0 + \Delta\beta$. $\Delta\beta$ is then determined by:

\be
\om (\beta_0 + \Delta\beta ) +\Delta\om = \om_0,\;
\om(\beta_0)=\om_0.
\ee
Here, $\om(\beta )$ is the relation between $\om$ and $\beta$ in the absence of the
Kerr effect, and we have neglected the change in $\Delta\om$ arising from the
shift in $\beta$, assuming that both $\Delta\om$ and $\Delta\beta$ are small.
We can then obtain $\Delta\beta$ by linear expansion:

\bea
\frac{\partial\om}{\partial\beta}\Delta\beta=-\Delta\om\Rightarrow
\Delta\beta=-\frac{\Delta\om}{v_g^0}
\eea
Here, $v_g^0$ is the waveguide group velocity in the absence of material
dispersion effects (since we consider propagation at a fixed frequency, the
frequency dependence of $\eps$ should not be taken into account when evaluating
$\frac{\partial\om}{\partial\beta}$). 
The change in $n_{\rm eff}$ arising from the Kerr effect is then:

\be
\Delta n_{\rm eff} = -\frac{c\Delta\om}{v_g^0\om_0}.
\ee

For $\Delta\eps_r$ we have:

\be
  \Delta\eps_r = 2\sqrt{\eps_r}n_2\mid {\bf E}\mid^2 + O\left(n_2^2\big|{\bf E}\big|^4\right),
\ee
and thereby, neglecting the small $n_2^2\mid {\bf E}\mid^4$ term and using
$\sqrt{\eps_r}=n_1$:

\be
\Delta n_{\rm eff} = \frac{ \eps_0c^3\int n_1n_2\mid {\bf E}\mid^4 dA}{v_g^0
\langle {\bf H,H}\rangle}=\frac{\eps_0 c\int n_1n_2\mid {\bf E}\mid^4 dA}{v_g^0
\langle {\bf E,D}\rangle} \label{dneff}
\ee
where, in the last equality, the inner product of {\bf H} with itself has been
rewritten using Eq. (\ref{feq}) and (\ref{lexampere}). The integral is taken over
the $xy$-plane. Note that, for a microstructured fiber, $n_1,n_2$ are 
position-dependent quantities.

  In the case of ordinary silica fibers, the parameters $n_1,n_2$ have little
variation over the fiber cross section, and it is common practice to express
the Kerr-induced change in the guided-mode effective index as:

\be
\Delta n_{\rm eff} = P\frac{n_2^P}{\maeff} \label{oldaeffdef}
\ee
where $P$ is the power launched into the fiber, $n_2^P=\frac{n_2}{n_1\eps_0 c}$
and \aeff is the effective mode area. In a microstructured fiber, both
$n_1$ and $n_2$ can have a strong position dependence, and a generalization of
Eq. (\ref{oldaeffdef}) is not straightforward. However, in the case where 
$n_1,n_2$ are piecewise constant functions, taking on $N$ different values over
the fiber cross section, we can modify the above definition to:

\be
\Delta n_{\rm eff} = P\sum_{i=1}^{N}\frac{n_{2i}^P}{A^i_{\rm eff}} \label{aeffdef}
\ee

where $n_{2i}^P$ denote the value of $n_{2}^P$ in the different sections of
the fiber.  It can be shown\cite{snyderandlove} that $P$ and $v_g^0$ are 
connected by:

\be
P=\int \left({\bf E\times H}\right)_z dA = v_g^0\langle {\bf E, D}\rangle \label{peq}
\ee

From Eqs. (\ref{dneff}),(\ref{peq}) we now obtain:

\be
\Delta n_{\rm eff} = P\sum_{i=1}^{N}n_{2i}^P\frac{(n_1^i\eps_0c)^2\int_i \mid {\bf E}\mid^4 dA}{\left(v_g^0\langle {\bf E, D}\rangle\right)^2}
=P\sum_{i=1}^{N}n_{2i}^P\frac{(n_1^in_g^0)^2\int_i \mid {\bf E}\mid^4dA}{\langle {\bf E, D}_r\rangle^2}\label{dneff-p}
\ee

In the last step, we have introduced ${\bf D}_r=\eps_r{\bf E}$ and the
effective group index of the guided mode, $n_g^0= \frac{c}{v_g^0}$. Note that
the integral over $\mid {\bf E}\mid^4$ in each term is restricted to the regions
with $n_1=n_1^i$. Comparing Eqs. (\ref{aeffdef}) and (\ref{dneff-p}), it can be seen that:

\bea
A^i_{\rm eff} = \frac{\langle {\bf E, D}_r\rangle^2}{(n_1^in_g^0)^2\int_i \mid {\bf E}\mid^4dA}\nonumber\\
=\left(\frac{n_1^i}{n_g^0}\right)^2
\frac{\langle {\bf E, D}_r\rangle^2}{\int_i \mid {\bf E\cdot D}_r\mid^2dA}  \label{aeffsum}
\eea

In the present paper, we shall only be concerned with the case of pure
silica/air fibers, in which $n_1,n_2$ are equal to 1 and 0, respectively,
in the air regions while having the values appropriate for silica in the
rest of the transverse plane. In this case, the nonlinear coefficient will
be entirely determined by the effective area relating to the silica regions, 
that is:

\be
\maeff = \left(\frac{n_{1}}{n_g^0}\right)^2
\frac{\langle {\bf E, D}_r\rangle^2}{\int_{\rm SiO_2} \mid {\bf E\cdot D}_r\mid^2dA}  \label{aeff}
\ee
The values for $n_1,n_2$ are now understood to be those of pure silica. 

At this point, two comments are in order: Firstly, note that of the two $n_1$ 
factors in the denominator of Eq. (\ref{aeff}), one comes from the definition
of $n_2^P$ in terms of $n_2$, whereas the other one comes from the fundamental
wave equation. Therefore, if one uses a table value of $n_2^P$ derived with a
$n_1$ different from that at which experiments are done, it may be necessary
to use two different $n_1$ values in the product. Secondly, Eq. (\ref{aeff})
differs somewhat from the formula commonly used for \aeff, namely:

\be
\tilde{A}_{\rm eff}=\frac{\left(\int \mid {\bf E}\mid^2 dA\right)^2}{\int \mid {\bf E}\mid^4 dA}\label{oldaeff}
\ee

However, if we assume that all the field energy resides in the silica regions 
of the fiber, and that $n_g^0\approx n_1$ we obtain:

\bea
\left(\frac{n_1}{n_g^0}\right)^2
\frac{\langle {\bf E, D}_r\rangle^2}{\int_{\rm SiO_2} \mid {\bf E\cdot D}_r\mid^2dA}=
\frac{\left(\int n_1^2\mid {\bf E}\mid^2 dA\right)^2}{(n_1n_g^0)^2\int\mid {\bf E}\mid^4 dA}
\approx\frac{\left(\int \mid {\bf E}\mid^2 dA\right)^2}{\int \mid {\bf E}\mid^4 dA} 
\eea

Thus, the commonly used formula (\ref{oldaeff}) appears as a limiting case of 
the more general result (\ref{aeff}). The approximations leading from Eq. 
(\ref{aeff}) to (\ref{oldaeff}) are reasonably well fulfilled in standard
fibers and in most index-guiding PCFs. This is, however, not the case 
for the fiber designs examined in the present work. In Fig.~\ref{fig2} we
show the relative difference between the \aeff definitions in Eqs.
(\ref{aeff}),(\ref{oldaeff}) defined as:

\be
\Delta \maeff=\maeff-\maefft
\ee
with \aeff, \aefft given by Eqs.  (\ref{aeff}),
(\ref{oldaeff}) respectively. The differences have been calculated for some
fibers designed to have a substantial part of the field energy in air, as 
discussed in the next section. It is evident that differences of 10-20\% 
between the two definitions can easily occur. Also, the difference varies
in both sign and magnitude over the transmission window of the fibers, so
that no simple scaling rule between the two definitions can be extracted.

  In this work, we solve Eq. (\ref{feq}) by expanding the magnetic field and
the dielectric function in plane waves, using a freely available software
package.\cite{johnson2001} Eq. (\ref{lexampere}) can be used to derive the electric field
vector from the magnetic, and the effective area can then be found from 
Eq. (\ref{aeff}). Since it is important for many applications of nonlinear
effects to be close to a wavelength at which the group velocity dispersion 
is zero, we have also investigated the dispersion properties of the fibers.
In order to scan a range of physical pitches efficiently we have used a 
recently developed perturbative approach to the inclusion of material 
dispersion effects\cite{laegsgaard}. We have found this scheme to be both
efficient and accurate in the case of silica/air microstructured fibers. 

\section{Numerical results and discussion}

In the present investigation, we will primarily focus on structures with 
$\frac{d}{\Lambda}$ lying in the interval between 0.3 and 0.8 for both core
and cladding holes.  Such structures are
by now routinely fabricated by fusing and drawing hand-stacked capillary tubes
and rods of silica. However, in order to investigate the limitations of the
hole-defect honeycomb PBG fibers, we have also studied a design with
\dclolns=0.95 and \dcol in the range between
0.1 and 0.3. Results for \dcolns=0.1, 0.2 and 0.3 are shown
in Fig.~\ref{fig3}. The lowest value of \aeff is 0.76$\lambda^2$ and is obtained
for \dcolns=0.2. The occurence of a minimum \aeff as a function of core
hole diameter can be understood as follows: Since nonlinear effects only
occur in the silica part of the fiber, reduction of core hole size increase
the region of integration in the denominator of Eq. (\ref{aeff}), thereby
acting to decrease the effective area. On the other hand, as is evident from
Fig.~\ref{fig3}, reduction of the core hole size also decreases the values of 
$\frac{\lambda}{\Lambda}$, where the guided mode becomes localized. In other
words, the fiber dimensions becomes larger relative to the wavelength of the 
guided mode, and this acts to increase the effective area, when measured 
relative to $\lambda^2$. 

   In the case of index-guiding PCFs, effective areas as low as 1.7 $\mu$m$^2$
at a wavelength of 1.55 $\mu$m have been reported experimentally, corresponding
to $\maeff\sim0.75\lambda^2$. The minimal mode area that can be obtained in
a silica rod in air has been proposed as a theoretical lower limit on the 
effective area in silica-based index-guiding fibers, and has been found to
be 1.48 $\mu$m$^2$ at 1.55 $\mu$m (or 0.62$\lambda^2$).\cite{birks1999} Thus, in spite of the
fact that the guided mode is localized around an airhole in the honeycomb
structures considered here they are able to obtain mode areas, which are only
slightly larger than what is possible in index-guiding PCFs.

  In Fig.~\ref{fig4} we show the effective area as a function of wavelength for 
$\frac{d_{cl}}{\Lambda}$=0.56, 0.68 and 0.80, and $\frac{d_{c}}{\Lambda}$ 
around 0.3-0.5. It can be seen that $\maeff\sim1.5-3\lambda^2$ is readily
obtained, and that the minimal area decreases with increasing size of the
cladding holes and increases for increasing size of the core hole. 

 In many experiments involving nonlinear effects it is important to work at
a wavelength at which the chromatic dispersion of the fiber is close to zero. 
In these situations, it is the minimal effective area 
obtainable at a given wavelength, under the condition that the dispersion
coefficient be zero, which is of interest. Both index- and PBG-guiding PCFs 
can have complicated dispersion curves with several zero-dispersion points.
 Some examples of what can be achieved with the 
honeycomb design for $\frac{d_{cl}}{\Lambda}$=0.68 are shown
in Fig.~\ref{fig5}. Following the three curves with $d_c/d_{cl}$=0.45 it can
be seen that the zero-dispersion point can have a discontinuous behaviour as
a function of $\Lambda$. The curve for $d_c/d_{cl}$=0.55 is an example of a
fiber with several dispersion zeros. To investigate the effective area at the
zero-dispersion point we have chosen to focus on the longest zero-dispersion
wavelength for a given design, since this is where one will usually have the
smallest effective area relative to the wavelength of the light. For the 
PBG fiber we have, therefore, investigated the
location of the longest zero-dispersion wavelength, $\lambda_0$  over a range of physical
fiber dimensions from $\sqrt{3}\Lambda$=1.5 to 5 $\mu$m. In Fig.~\ref{fig6} we
report $\lambda_0$ versus the physical pitch, $\Lambda$, and in Fig.~\ref{fig7} the 
effective area at $\lambda_0$ is plotted versus $\lambda_0$. Broken curves indicate
a discontinuity in $\lambda_0$ as a function of $\Lambda$. In some cases, the
zero-dispersion wavelength can sweep over the same frequency several times as
$\Lambda$ is varied, which is why some of the curves in Fig.~\ref{fig7} are 
multi-valued.  Generally, the honeycomb fibers 
tend to have dispersion zeros falling in the range between 0.8 and 2 $\mu$m.
 
For comparison we have also considered index-guiding PCFs with a triangular array
of airholes constituting the cladding and a core-defect formed by a missing air hole. 
For air hole diameters $d\leq d^*$ (with $d^* \sim 0.45 \Lambda$) this class of PCFs 
is endlessly single mode\cite{knight1996} and by scaling the pitch both large-mode area 
PCFs as well as small-core non-linear PCFs can be formed. In addition to the mode-size 
the dispersion properties may also be engineered. In the top panel of Fig.~\ref{fig8} 
we show the zero-dispersion wavelength $\lambda_0$ as a function of the pitch $\Lambda$ 
for 5 hole diameters. Depending on the pitch the PCF may have none, one, or two (or 
even three if also considering the near-infrared regime) dispersion zeros. For the 
situation with two dispersion zeros it is seen that the second dispersion zero 
(counting from the short wavelength limit) depends strongly on both the pitch and the 
hole diameter whereas for the first dispersion zero the dependence on hole diameter 
dominates over the much weaker dependence on pitch. One of the exciting properties of 
these PCFs is that by increasing the hole diameter the lowest dispersion zero can be 
moved toward the visible regime. In the lower panel of Fig.~\ref{fig8} we show the 
effective area versus zero-dispersion wavelength. In general we find that when 
the hole size is increased the mode becomes more tightly confined with a 
smaller effective area. The plot also illustrates a highly desired property; 
when shifting the first dispersion-zero toward the visible the effective 
area also decreases so that the intensity thresholds for various non-linear 
phenomena also decreases. We note that the first dispersion zero may be 
moved to the visible and the effective area may be decreased by further 
increasing the air hole diameter (so that $d >d^*$), but then care must 
be taken that the PCF remain single-mode near the dispersion zero.  
 
  Comparing Figs.~\ref{fig7} and ~\ref{fig8} it can be seen that the honeycomb PBG
fibers investigated here do not offer substantially improved flexibility in the
tailoring of mode areas and dispersion properties compared to the index-guiding PCFs. 
The PBG designs with large cladding holes do seem to offer smaller mode areas at the
longer values of $\lambda_0$ but this could also be obtained in index-guiding PCFs by
increasing the size of the airholes, thereby going further out of the endlessly 
single-mode regime. Determining the single-mode
regime for PBG fibers with large cladding airholes is difficult due to the appearance
of multiple bandgaps even at relatively long values of $\frac{\lambda}{\Lambda}$. 
For the fibers with  $\frac{d_{cl}}{\Lambda}$=0.8 we have found that a guided
second-order mode is present at $\lambda_0$ over most of the $\Lambda$ range
investigated. For the fibers with smaller values of  $\frac{d_{cl}}{\Lambda}$ the
second-order modes mostly appears at wavelengths shorter than $\lambda_0$ for 
$\Lambda$-values smaller than  $\sim$2 $\mu$m meaning that a useful range of $\lambda_0$
values without second-order mode guidance is available. We have not, however, 
checked for the presence of guided modes in higher-order gaps.

  One interesting feature of the honeycomb PBG design compared to the
index-guiding PCFs is that a relatively large fraction of the field energy
resides in the airholes of the fiber. This is illustrated in Fig.~\ref{fig9}, where
the fraction of electric field energy present in air has been plotted for
the fiber designs discussed above. It is evident that energy fractions of
10-15\% in air are readily obtained, even for holes of moderate size, and that 
the fraction increases with increasing size of the core hole defect. 
To estimate the
range of energy distributions accessible, we have investigated some designs,
in which the core hole defect has been further enlarged. In Fig.~\ref{fig10}, we show
results for effective areas and the energy fraction in air for fibers with
$\frac{d_c}{\Lambda}$ around 0.6 and varying size of the cladding holes.
It can be seen that the energy fraction in air can be as large as 30\%, while
still having a range of possible zero-dispersion wavelengths and fairly
small mode areas. Further increase of $d_c$ does not push appreciably more
field energy into the airholes, however, the transmission windows and
accessible zero-dispersion ranges are quickly diminished. The somewhat counterintuitive
fact that the that the minimum of the effective area curves (i.e., the maximum of the
nonlinear coefficient at a particular wavelength) occurs approximately at
the same $\frac{\lambda}{\Lambda}$ values as the maximum of the energy fraction
in the airholes is due to the fact that the energy fraction in
air is high at long wavelengths, where the fiber size relative to wavelength
is small.

 In Fig.~\ref{fig11}
we show some radial mode profiles obtained by integrating the electric field
energy density over the angular coordinate in a polar coordinate system around
the core center. The curves are calculated for fibers with \dclol=0.8 and varying
size of the core hole, and have been normalized so that their radial integral
is unity. The wavelength has been chosen so as to maximize the fraction of field 
energy in air. Only 5-10\% of the field energy is present in the central hole
defining the core. This shows that the increase in the energy fraction present
in air for increasing \dcol is not so much due to the field energy being pushed
into the central core hole, but rather to an increase of the field energy 
present in the cladding region as is also evident from Fig.~\ref{fig11}.
The presence of a substantial part of the field energy in air not only influences
the integrals in Eq. (\ref{aeff}), but also makes the group velocity of the guided
mode deviate substantially from the material refractive index, thus influencing
the prefactor $\left(\frac{n_1}{n^0_g}\right)^2$. In Fig.~\ref{fig12}, this fraction is depicted for the 
designs with $\frac{d_{cl}}{\Lambda}$=0.8. Its importance for a correct
evaluation of the effective area in these structures is evident.

The power fraction in air (which is not completely equivalent to the quantity
calculated here) for index-guiding PCFs was investigated by Monro and
co-workers\cite{monro-evf} who found that large airhole diameters and $\frac{\lambda}{\Lambda}$-values 
were needed to obtain appreciable power fractions in air. In Fig.~\ref{fig13}, 
we report the fraction of field energy in air for some index-guiding PCFs with
the same cladding structure as those discussed above but somewhat larger 
airholes. Comparing Figs.~\ref{fig9} and ~\ref{fig10} with Fig.~\ref{fig13} it is
evident that the $\lambda/\Lambda$ values needed to push a given fraction of
field energy into the air region are considerably larger in the index-guiding
fibers, calling for smaller values of the physical pitch for operation at a
given wavelength.  This may make the honeycomb fibers particularly relevant 
for evanescent field devices,\cite{monro-evf-2001,hoo-evf} such as gas sensors, based on interactions with 
the glass, through, e.g., nonlinear effects or inscribed Bragg gratings. 

\section{Conclusion}

 In conclusion, we have investigated the field energy distribution and nonlinear 
coefficients of honeycomb photonic bandgap fibers and compared them to 
index-guiding photonic crystal fibers with a cladding structure consisting
of a triangular array of airholes. A generalized concept of effective mode
area, which is adequate for the treatment of fibers with a substantial part of
the field energy present in the airholes, has been derived for this purpose.
While the honeycomb fibers do not seem to offer increased flexibility in 
the design of dispersion properties and mode areas they do offer
the same possibilities as the index-guiding fibers, at wavelengths
above $\sim$1 $\mu$m. In addition, the honeycomb fibers have a larger fraction
of the field energy present in the airholes which may make these fibers 
particularly interesting for sensor applications based on interactions with
the glass, through e.g. nonlinear effects and/or inscribed Bragg gratings.


\newpage

\section*{List of Figure Captions}

\begin{figure}[h]
\resizebox{13cm}{!}{\includegraphics[0cm,0cm][20cm,29cm]{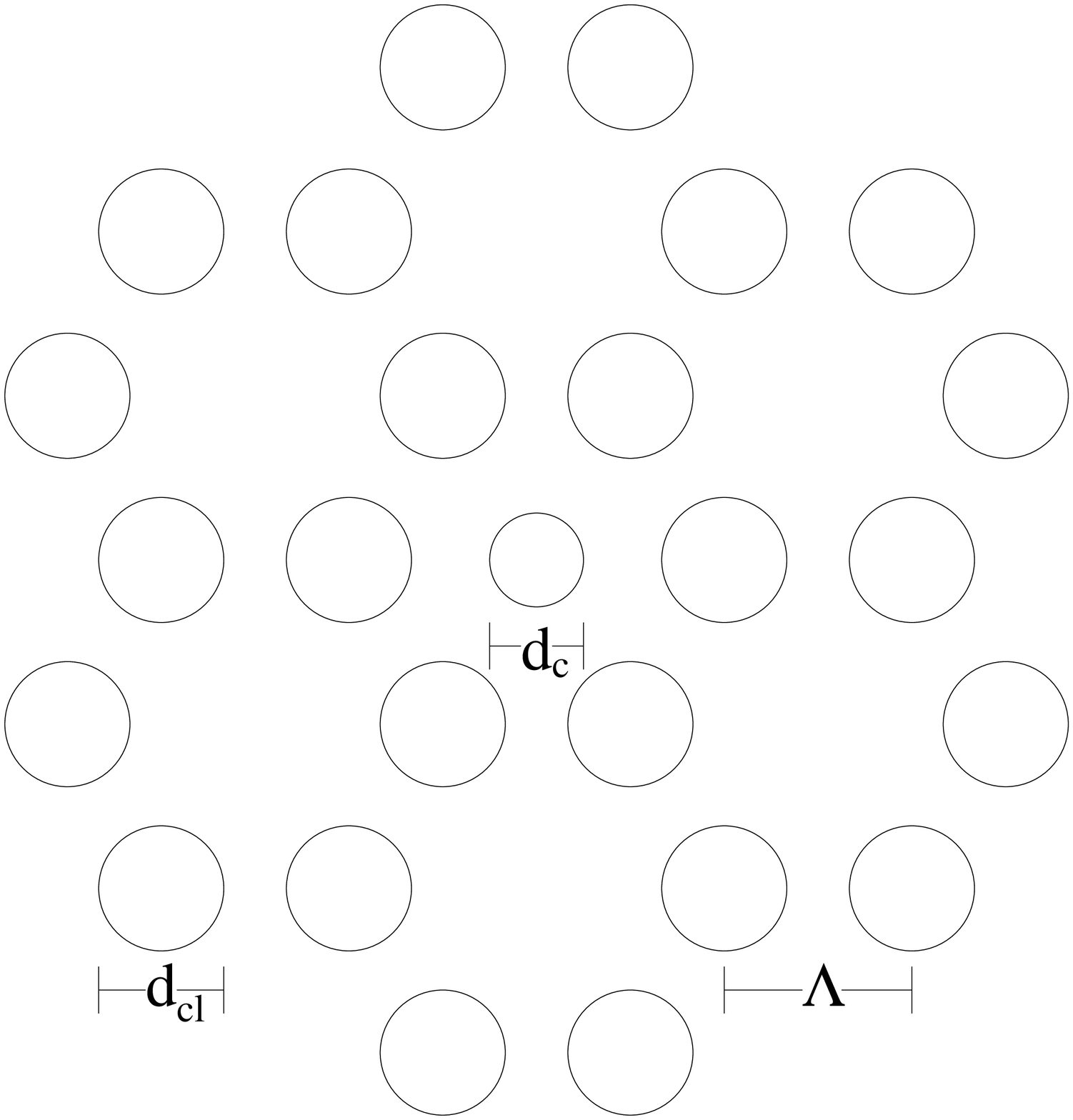}}
\caption{L{\AE}GSGAARD}
\label{fig1}
\end{figure}

Fig.~\ref{fig1} The generic PCF structure investigated in the present work. The core
and innermost cladding holes are shown along with the defining parameters
$d_c$, $d_{cl}$, and $\Lambda$.

Fig.~\ref{fig2} Relative difference between the effective area definition
proposed here, Eq. (\ref{aeff}), and the commonly used definition, 
Eq. (\ref{oldaeff}).

Fig.~\ref{fig3} Effective area, relative to wavelength, calculated from Eq. (\ref{aeff})
for a fiber with $\frac{d_{cl}}{\Lambda}$=0.95 and various values of the core
hole diameter, $d_c$. 

Fig.~\ref{fig4} Effective area relative to the wavelength of the guided mode for 
honeycomb PBG fibers with $\frac{d_{cl}}{\Lambda}$=0.56 (a), 0.68 (b) and
0.8 (c) and various values of $\frac{d_{c}}{d_{cl}}$.

Fig.~\ref{fig5} Plots of the chromatic dispersion coefficient, D, in units of ps/nm/km
for various values of the pitch in a honeycomb PBG fiber with $\frac{d_{cl}}{\Lambda}$=0.68
and $\frac{d_c}{d_{cl}}$=0.45.

Fig.~\ref{fig6} Longest zero-dispersion wavelength, $\lambda_0$, as a
function of the physical pitch, $\Lambda$. Structures and labeling as in Fig.~\ref{fig4}.

Fig.~\ref{fig7} Effective area at the zero-dispersion wavelength, $\lambda_0$ as a
function of $\lambda_0$. Structures and labeling as in Fig.~\ref{fig4}.

Fig.~\ref{fig8} Effective area and dispersion zeros for index-guiding PCFs. Top panel: 
Zero-dispersion wavelength versus pitch. Lower panel: Effective area at the 
zero-dispersion wavelength as a function of the zero-dispersion wavelength.

Fig.~\ref{fig9} The fraction of the electric field energy of the guided mode present
in the airholes. Structures and labeling as in Fig.~\ref{fig4}.

Fig.~\ref{fig10} Effective area relative to wavelength (a), effective area at the
zero-dispersion wavelength (b) and energy fraction in air (c) for some fiber
designs in which a substantial part of the field energy resides in the 
airholes.

Fig.~\ref{fig11} Radial profile of the electric field energy density, obtained 
by integration over the angular coordinate in a coordinate system with origin
at the core center. The curves are normalized to have unit radial integrals.
The thin vertical lines indicate the position of the first ring of cladding
airholes.

Fig.~\ref{fig12} Variation of the prefactor $\frac{n_1}{n^0_g}$ in Eq. (\ref{aeff})
with wavelength for a fiber with $\frac{d_{cl}}{\Lambda}$=0.8 and various 
values of the core hole diameter, $d_c$.

Fig.~\ref{fig13} Energy fraction in air of index guiding fibers with a cladding 
structure as shown in Fig.~\ref{fig8} and various airhole diameters $d$.

\newpage

\begin{figure}[h]
\resizebox{13cm}{!}{\includegraphics[0cm,0cm][20cm,29cm]{fig2.eps}}
\caption{L{\AE}GSGAARD}
\label{fig2}
\end{figure}

\begin{figure}[h]
\resizebox{13cm}{!}{\includegraphics[0cm,0cm][20cm,29cm]{fig3.eps}}
\caption{L{\AE}GSGAARD}
\label{fig3}
\end{figure}

\begin{figure}[h]
\resizebox{13cm}{!}{\includegraphics[0cm,0cm][20cm,29cm]{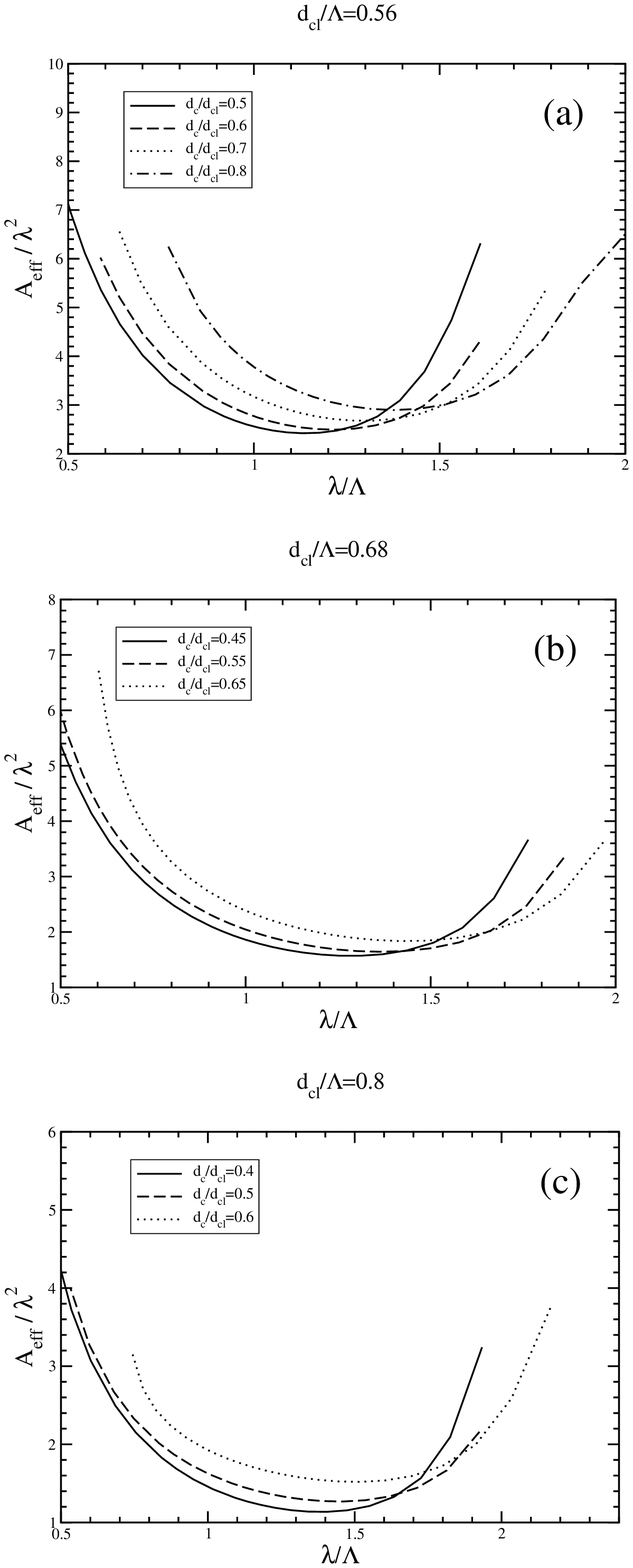}}
\caption{L{\AE}GSGAARD}
\label{fig4}
\end{figure}

\begin{figure}[h]
\resizebox{13cm}{!}{\includegraphics[0cm,0cm][20cm,29cm]{fig5.eps}}
\caption{L{\AE}GSGAARD}
\label{fig5}
\end{figure}

\begin{figure}[h]
\resizebox{13cm}{!}{\includegraphics[0cm,0cm][20cm,29cm]{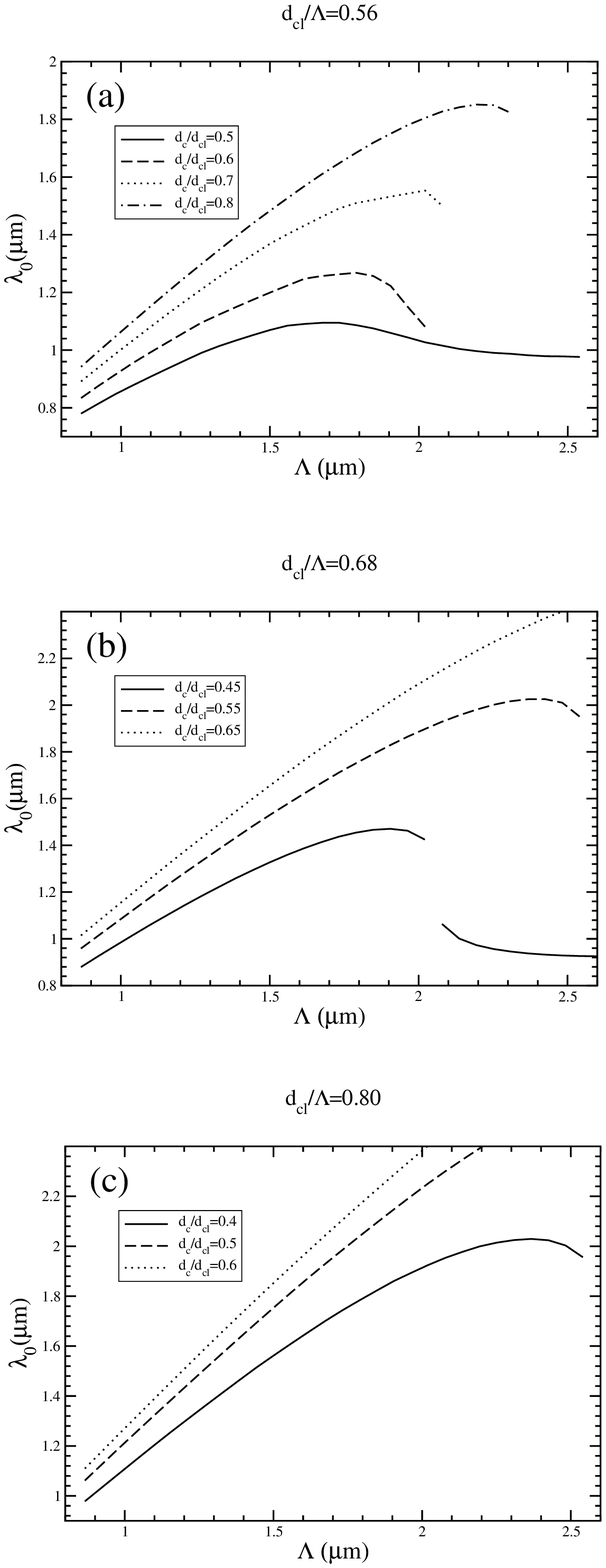}}
\caption{ L{\AE}GSGAARD}
\label{fig6}
\end{figure}

\begin{figure}[h]
\resizebox{13cm}{!}{\includegraphics[0cm,0cm][20cm,29cm]{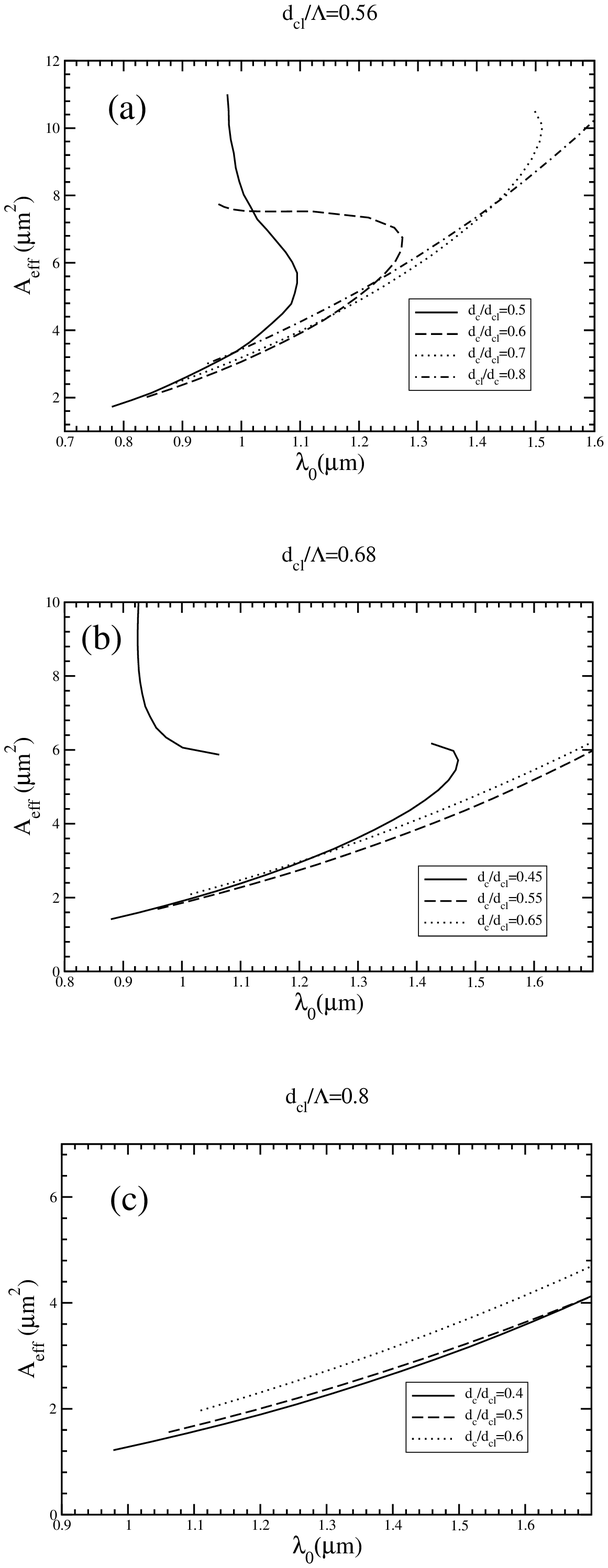}}
\caption{L{\AE}GSGAARD}
\label{fig7}
\end{figure}

\begin{figure}[h]
\resizebox{13cm}{!}{\includegraphics[0cm,0cm][20cm,29cm]{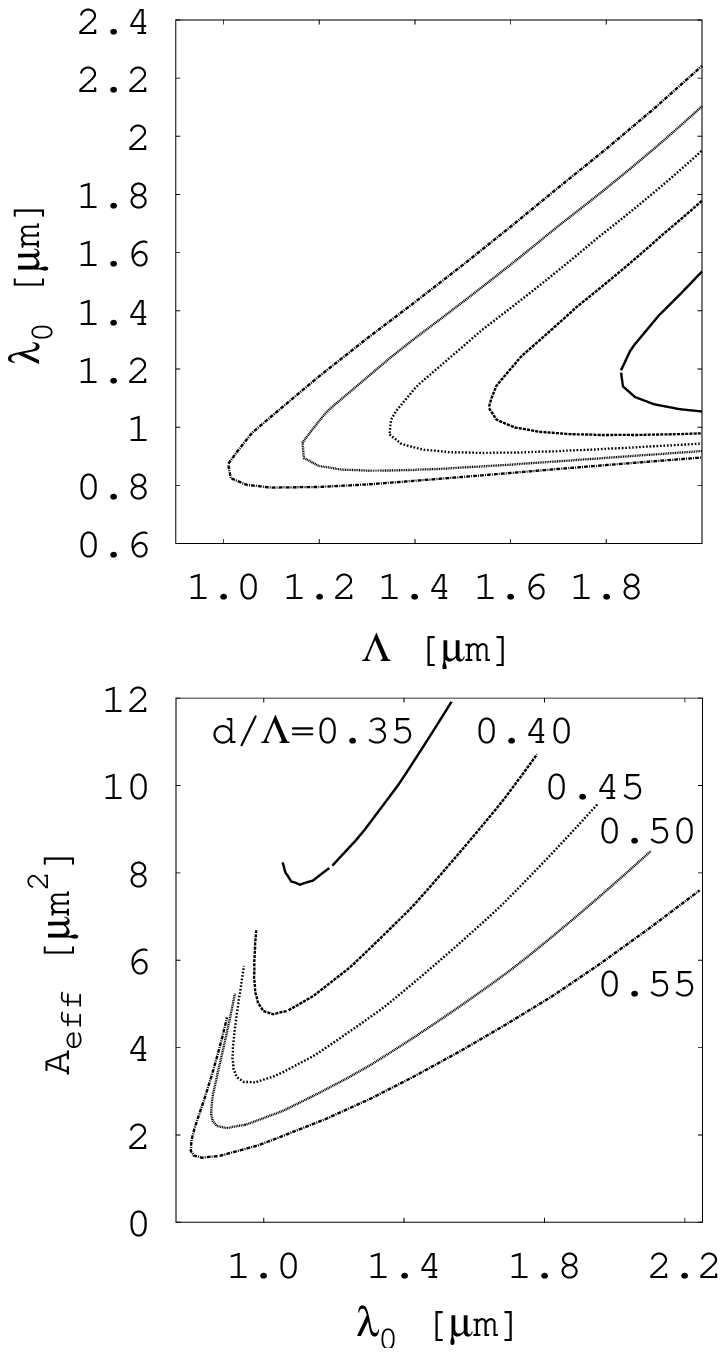}}
\caption{L{\AE}GSGAARD}
\label{fig8}
\end{figure}

\begin{figure}[h]
\resizebox{13cm}{!}{\includegraphics[0cm,0cm][20cm,29cm]{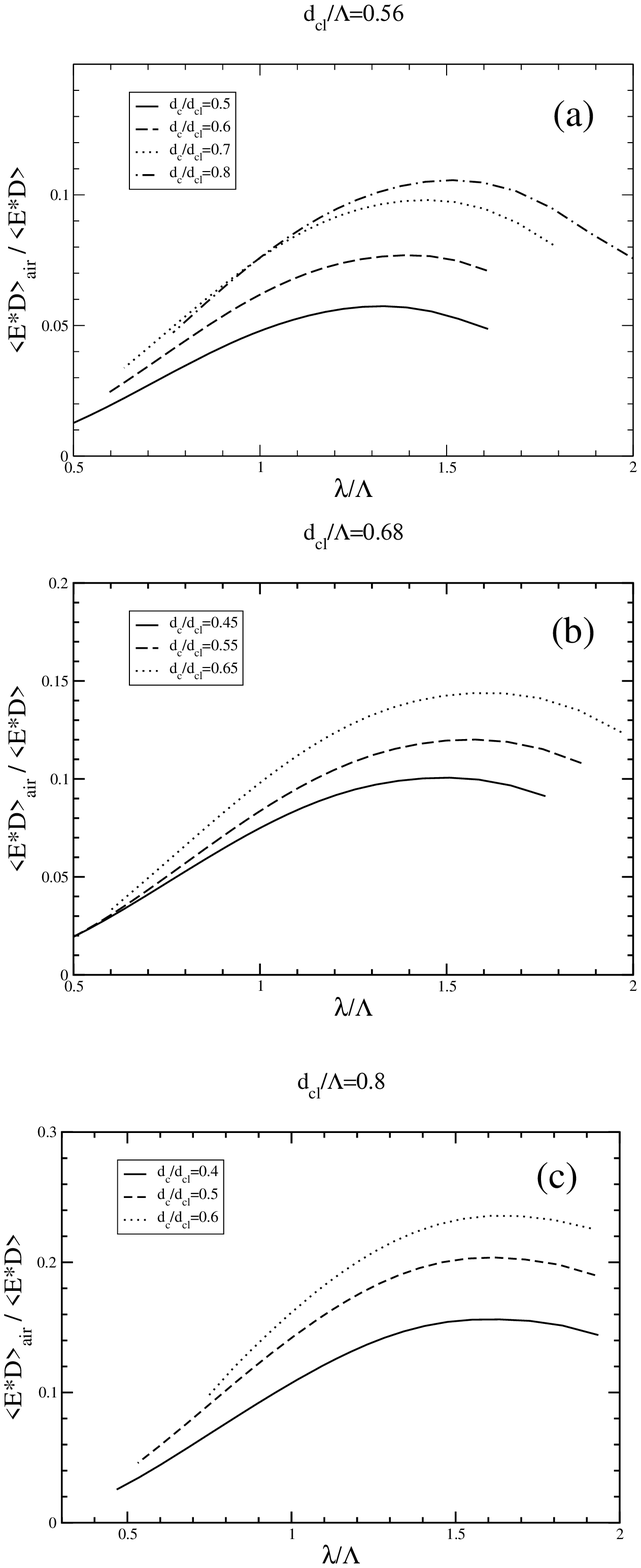}}
\caption{L{\AE}GSGAARD}
\label{fig9}
\end{figure}

\begin{figure}[h]
\resizebox{13cm}{!}{\includegraphics[0cm,0cm][20cm,29cm]{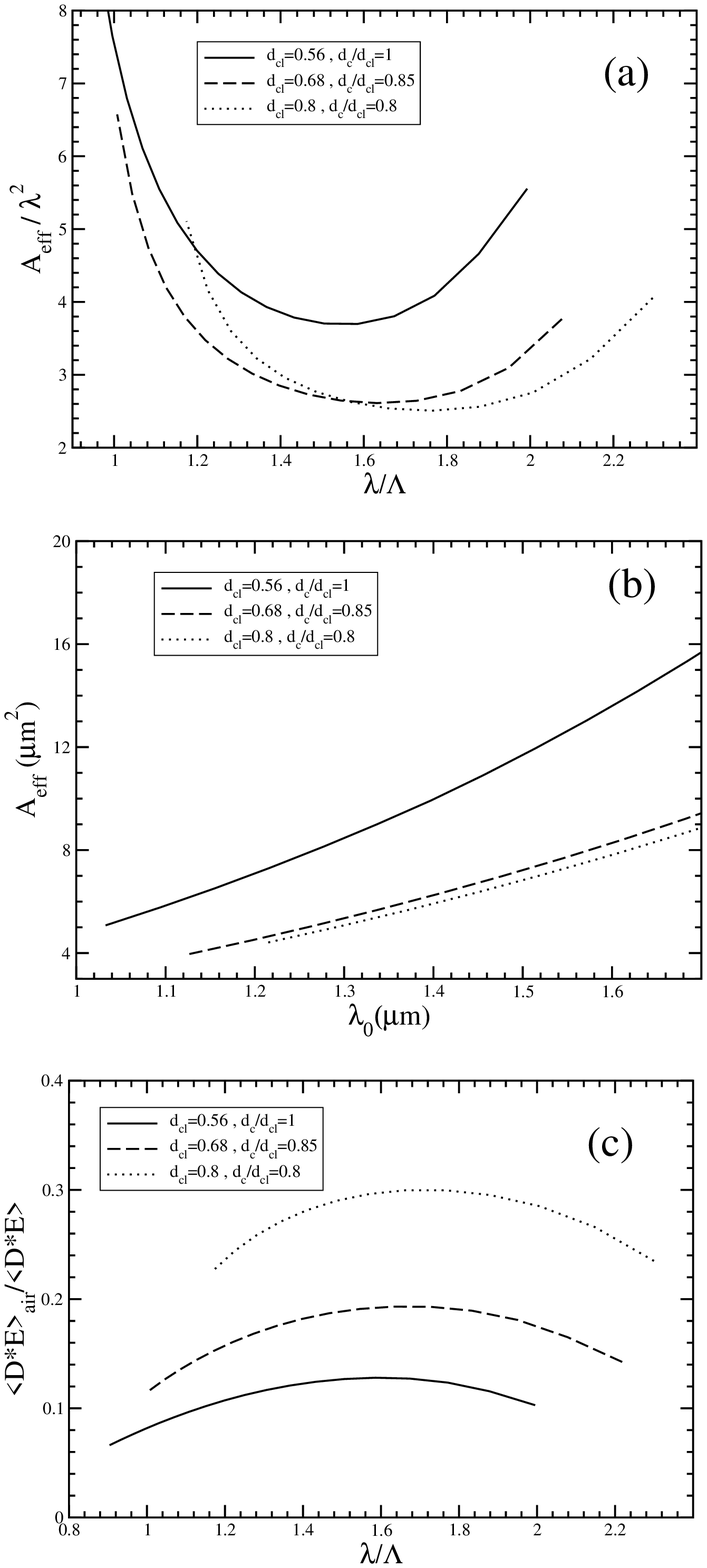}}
\caption{L{\AE}GSGAARD}
\label{fig10}
\end{figure}

\begin{figure}[h]
\resizebox{13cm}{!}{\includegraphics[0cm,0cm][20cm,29cm]{fig11.eps}}
\caption{L{\AE}GSGAARD}
\label{fig11}
\end{figure}

\begin{figure}[h]
\resizebox{13cm}{!}{\includegraphics[0cm,0cm][20cm,29cm]{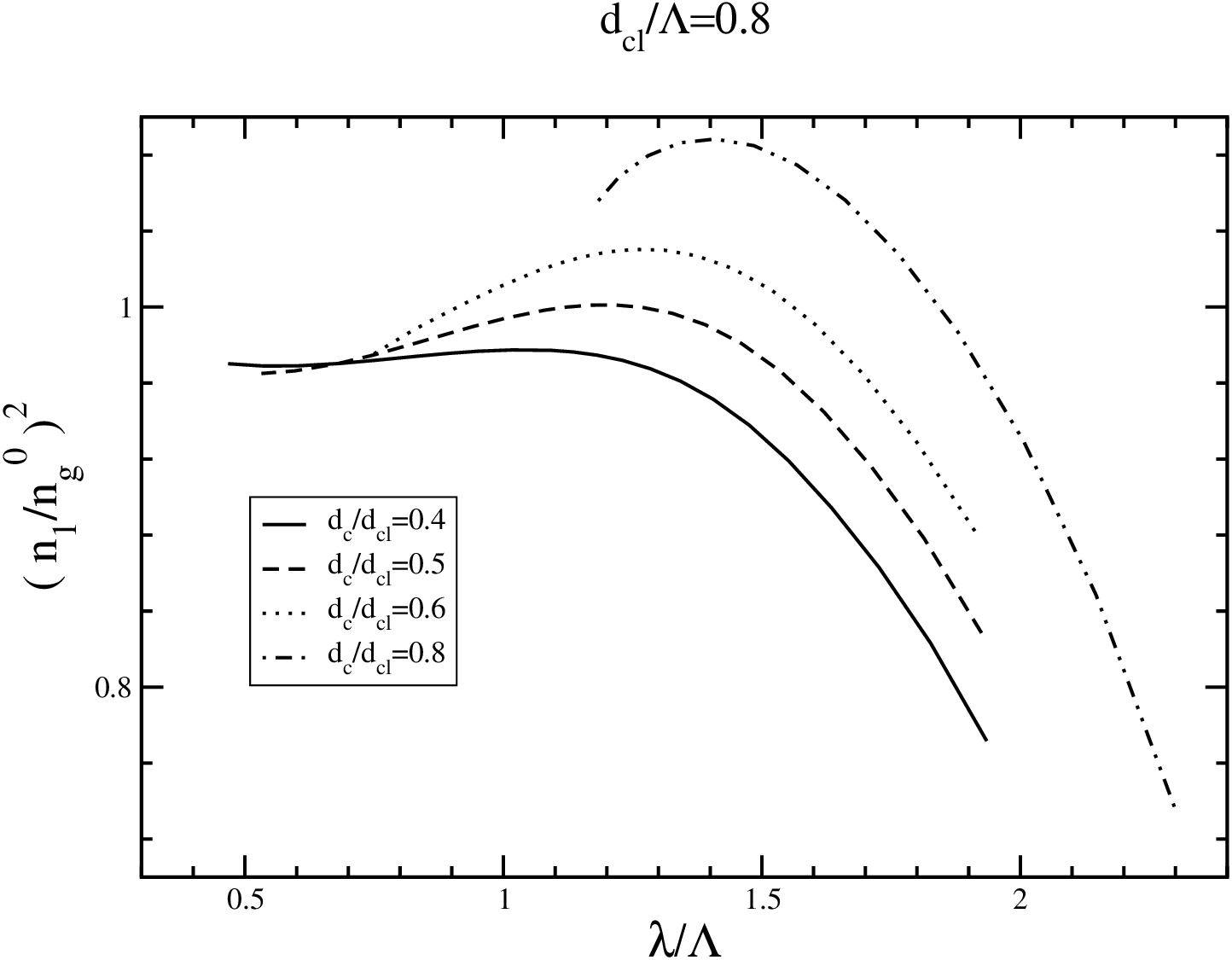}}
\caption{L{\AE}GSGAARD}
\label{fig12}
\end{figure}

\begin{figure}[h]
\resizebox{13cm}{!}{\includegraphics[0cm,0cm][20cm,29cm]{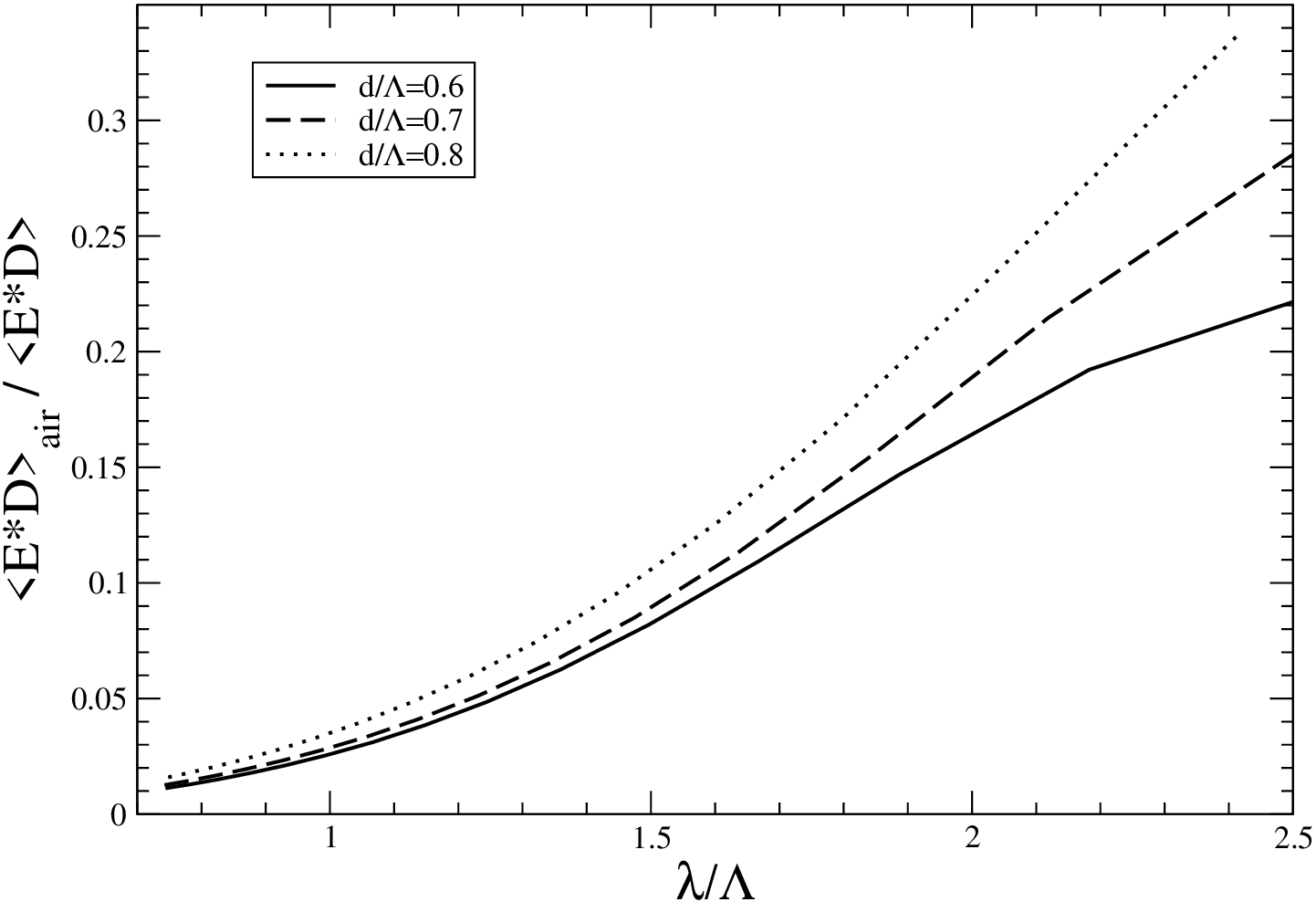}}
\caption{L{\AE}GSGAARD}
\label{fig13}
\end{figure}

\end{document}